\documentclass[]{aa}
\usepackage{txfonts}
\usepackage{graphicx}
\usepackage{natbib}
\bibpunct{(}{)}{;}{a}{}{,} 

\def\Esn{E_{\rm SN}}
\def\Ed{{\cal E}}

\begin{document}

\title{On the plasma temperature in supernova remnants with cosmic-ray modified shocks}

\author{L. O'C. Drury \inst{1}
\and F. A.  Aharonian \inst{1,2}
\and D. Malyshev \inst{1}
\and S. Gabici \inst{1}}

\institute{Dublin Institute for Advanced Studies, 31 Fitzwilliam Place, Dublin 2, Ireland; \and Max-Planck-Institut f\"ur Kernphysik, Saupfercheckweg 1, 69117 Heidelberg, Germany}

\date{}

\abstract{Multiwavelength observations of supernova remnants  can be explained within the framework of 
the diffusive shock acceleration theory,  which allows effective conversion of the explosion energy 
into cosmic rays.  Although the models of nonlinear shocks describe reasonably well 
the nonthermal  component of emission,  certain issues, including the heating of the thermal 
plasma and the related X-ray emission, remain still open. }
{To discuss how the evolution and structure of supernova remnants is affected by strong particle acceleration at the forward shock.}
{Analytical estimates combined with detailed discussion of the physical processes.}{The overall dynamics is shown to be relatively insensitive to the amount of particle acceleration, but the post-shock gas temperature can be reduced by the acceleration to a multiple, even as small as six times, the ambient temperature with a weak dependence on the shock speed.  This is in marked contrast to models with no particle acceleration where the post-shock temperature is insensitive to the ambient temperature and is determined by the square of the shock speed. It thus appears to be possible to suppress effectively 
thermal X-ray emission from remnants by strong particle acceleration. This might provide a clue for understanding the 
lack of thermal X-rays from the TeV bright supernova remnant RX J1713.7-3946.}{}


\maketitle

\section{Introduction}

The purpose of this article is to discuss  how the evolution of supernova remnants may be modified by strong particle acceleration at the forward shock and the resulting implications for observational diagnostics, in particular those using X-rays. This is an interesting problem in its own right, but is becoming of increasing importance because of the rapidly improving observational situation; there are now a number of shell-type SNRs detected as sources of TeV 
gamma rays (see for a recent review \citealt{Gabici} and 
references therein), which certainly proves the acceleration of particles to energies of at least $10^{14}\,\rm eV$, in 
particular in   SNR  RX J1713.7-3946 \citep{HESS_last},  but the interpretation of the observations remains controversial, in part because of the neglect or misunderstanding of some of these effects.  In particular, recently a strong claim has been made against the hadronic origin of the  TeV-gamma-ray emission from  the young SNR  RX J1713.7-3946 \citep{Katz},  
based on the lack of thermal X-rays from this 
TeV-bright object  \citep{Slane,XMM,Suzaku,Tanaka}. However, given the complex nature of the problem, which involves the 
physics of nonlinear shocks as well uncertainties  related to the heating mechanism of electrons, such a claim is premature and further studies are needed.    
In this paper we discuss the impact of the efficient particle acceleration in strong nonlinear 
shocks on the heating of the ion component of plasma.   
Our intent is to give a physically motivated account of the key processes.  We feel that this is more important at this stage than the construction of complicated numerical models, although these will clearly be required for the detailed understanding of specific sources.

\section{Dynamics of supernova remnants}

In essence a supernova remnant, once it has expanded out of the initial free-expansion phase where the total energy is dominated by the kinetic energy of the expanding ejecta from the explosion of the progenitor star, is just a hot bubble of high-pressure gas pushing a shell of swept-up material into a low pressure external medium.
It is important to realise that this is a dynamic situation and that even if the system appears to evolve in a quasi-steady manner (as in the generalised Sedov solutions) this represents a dynamical balance between the conversion of internal pressure into kinetic energy of expansion by  ``$PdV$''  work on the one hand and the competing conversion of bulk kinetic energy into random particle motion, i.e.\ pressure, in the shock on the other.  

If the radius of the bubble is $R(t)$ and the explosion energy is $\Esn \approx 10^{51}\,\rm erg$ then to within factors of order unity the total pressure within the bubble has to be $3\Esn/4\pi R^3$ with only a weak dependence on whether this pressure comes from cosmic rays, thermal gas or magnetic fields.  More precisely, for relativistic accelerated particles and tangled magnetic fields the pressure is $1/3$ the energy density, and for non-relativistic thermal particles the pressure is $2/3$ the energy density.  The average interior pressure is thus between
$1/3$ and $2/3$ of $3(\Esn-E_{\rm K})/4\pi R^3$ where $E_{\rm K}$ is the kinetic energy of the remnant expansion, which we can take as some roughly constant fraction of the total energy $\Esn$. Equating this internal pressure, again within factors of order unity, to the ram pressure of the external medium flowing into the shock gives the expansion of the remnant, namely
\begin{equation}
\rho_0 \left(d R\over d t\right)^2 \approx {\Esn\over R^3}
\end{equation}
or
\begin{equation}
{d R\over d t} \approx R^{-3/2} \left(\Esn\over\rho_0\right)^{1/2}
\end{equation}
from which the well-know Sedov self-similar relation,
\begin{equation}
R(t) \approx \left(\Esn\over\rho_0\right)^{1/5} t^{2/5}
\end{equation}
immediately follows.  
The shock velocity $V$ then scales as
\begin{equation}
V = {d R\over d t} ={2\over 5} {R\over t} \propto t^{-3/5} \propto R^{-3/2}.
\end{equation}

It is worth noting that the factors of order unity mentioned above all go with the factor $\Esn/\rho_0$ and thus affect the final result for the radius as a function of time only to the very weak power $1/5$.  It follows immediately that particle acceleration has a relatively small effect on the bulk expansion of the remnant and that the basic Sedov scaling for the shock radius as a function of time, explosion energy and external density is very robust. However the shock structure itself, and the internal structure of the remnant, are much more sensitive to the effects of particle acceleration.

The primary effect of putting energy into particle acceleration is to reduce the amount of energy available for gas heating \citep{don,giulia}, but it also leads to increased shock compression ratios.  In part this is due to the softer equation of state associated with the relativistic particles, but the more significant reason is the need to constantly pump  energy into particles near the maximum energy (upper cut-off) of the spectrum.  There are three distinct reasons for this.
\begin{itemize}
\item (i) One is to compensate for high-energy particles physically escaping from the system.  This is clearly important if the effective magnetic field, and thus the upper cut-off energy for the spectrum of accelerated particles, is a decreasing function of  time as is the case with the currently fashionable field amplification models (see e.g., \citealt{bell}).   The effect has perhaps been somewhat over-emphasised by the wide-spread use of so-called ``free escape boundaries" as technical devices to regularise steady planar models of shock acceleration (see, e.g., \citealt{ellison85, RKD}).

\item (ii) The second, often overlooked, reason is to compensate for the geometrical expansion of the acceleration region as the shock evolves.    In an expanding and decelerating spherical shock the acceleration volume scales as the surface area of the shock times the diffusion length-scale, both increasing functions of time, and thus an ever larger number of particles are needed at a given energy to maintain the particle intensity.  At energies well below the cut-off the acceleration time-scales are very short compared to the dynamical evolution times and the system can easily stay in equilibrium, but as we approach the upper cut-off energy the time-scales become comparable and this dilution effect becomes a significant drain on the available supply of accelerated particles.  In fact, if no other effects (radiative losses, plasma processes, collisions etc.) intervene, it is precisely the on-set of this dilution that determines the location of the upper spectral cut-off.

\item (iii) Thirdly, as we will see, in the approximately self-similar Sedov-like phase of the evolution the interior pressure in the remnant is only about half the immediate post-shock value.  There is thus a gradient in the accelerated particle pressure in the down-stream post-shock region which must be associated with a diffusive flux from behind the shock into the interior of the remnant \citep{DMV}. This effect is entirely absent in steady planar solutions.

\end{itemize}\noindent
Compared to a stationary planar shock solution, these effects all represent additional energy sinks (from the shock itself; only the first is a real energy loss from the whole system) and just as in the case of radiatively cooled gas shocks lead to increased shock compression ratios (typically 10 or more is seen in simulations).  This is the fundamental reason why steady one-dimensional shock models, such as the original two-fluid models \citep{DV81} and their descendants,  are of limited applicability to supernova remnants; in addition to the energy loss by advection downstream, there has to be a significant flux of energy to particles at the upper cut-off $\Phi$
and thus the mass, momentum and energy conservation relations for the shock take the approximate modified form
\begin{eqnarray}
\label{eqs-boundary-modified-1}
\rho_0 U_0 &=& \rho_1 U_1 = A\\ \nonumber
\rho_0 U_0^2 &=& \rho_1 U_1^2 + P\\ \nonumber
{1\over 2}\rho_0 U_0^3 &=& {1\over 2}\rho_1 U_1^3 + U_1\left(\Ed+P\right)+ \Phi \nonumber
\end{eqnarray}
where $U_0$ is the upstream velocity, $U_1$ the post-shock decelerated velocity, $\Ed$ is the total post-shock energy density of thermal and non-thermal particles, $P$ the associated total pressure and we further assume that these are negligible upstream of the shock. It is important to note that these relations are only approximate for an expanding spherical system because the shock is not planar and is not in a stationary equilibrium, but the additional energy flux term captures much of this non-planar and non-equilibrium character.  These imply
\begin{eqnarray}
A \left(U_0-U_1\right) &=& P\\
{1\over 2} P \left(U_0+U_1\right)& =&  U_1\left(\Ed + P\right)+ \Phi
\end{eqnarray}
and the latter implies that the shock compression ratio is given by
\begin{equation}
s= {U_0\over U_1} =  1+ {2 \Ed\over P} + {2\Phi\over U_1 P}.
\end{equation}
For non-relativistic particles $\Ed/P=3/2$ rising to $\Ed/P=3$ for relativistic particles;
thus $1+ 2\Ed/P$ lies between 4 and 7 with an additional positive contribution from the $2\Phi/U_1 P$ term.

We can estimate the order of magnitude of the $\Phi$ term as follows.  From basic shock acceleration theory (see Appendix A) the number flux of particles being accelerated upwards in energy at momentum $p$ is
\begin{equation}
{4\pi p^3\over 3} f(p) \left(U_0 - U_1\right)
\end{equation}
and thus the associated energy flux at the upper cut-off is
\begin{equation}
{4\pi p_{max}^3\over 3} f(p_{max}) \left(U_0 - U_1\right) c p_{max}.
\end{equation}
But this is essentially just the integrand in the total non-thermal particle pressure when expressed as an integral over logarithmic momentum,
\begin{equation}
P_C = \int {p v\over 3} 4\pi p^2 f(p)\, dp = \int {4\pi p^3\over 3} pv f(p) d \ln(p)
\end{equation}
and thus, if the pressure is contributed more or less uniformly per logarithmic interval as in the test-particle case, $f(p)\propto p^{-4}$, we have that
\begin{equation}
\Phi \approx {P_C (U_0 - U_1)\over\lambda},
\end{equation}
where $\lambda>1$ is the ratio of the total non-thermal pressure to that part contributed by particles at the upper cut-off.  For an equal energy per logarithmic interval spectrum,$f(p)\propto p^{-4}$, it is simply the logarithmic range of the spectrum,  $\lambda =\ln(p_{\rm max}/mc)$ dropping to $\lambda = 2$  for Malkov's universal strongly modified shock spectrum, $f(p)\propto p^{-3.5}$, where most of the energy is carried near the upper cut-off \citep{malkov}.
It is then easy to see that the shock compression is given by
\begin{equation}
s = 1 + {2\Ed\over P - 2 P_C/\lambda},
\end{equation}
which can easily be quite large for significant particle acceleration and small values of $\lambda$.    Writing $P_G$ for the thermal pressure and $P_C$ for the non-thermal pressure and assuming the latter to be dominated by relativistic particles we have
\begin{equation}
s = 1 +  {3P_G + 6 P_C\over P_G + P_C\left(1-2/\lambda\right)}.
\end{equation}
It is very remarkable that even at this crude macroscopic level we can see a connection between extreme acceleration efficiency with the shock compression tending to infinity and a spectral hardening towards an exponent of $3.5$ corresponding to $\lambda = 2$
as required by Malkov's asymptotic solution.

This large compression ratio allows us to retrospectively justify the simplified relations we started with using an approximation originally due to \citet{Chernyi} (see also \citealp{Zeld}).   The key is to assume that all the swept-up matter is concentrated in a thin shell immediately behind the shock with radial velocity $U_0-U_1$ and total mass
\begin{equation}
M= {4\pi \over 3}R^3 \rho_0.
\end{equation}
If the pressure in the interior of the remnant is $P_{\rm int}$, then Newton's law of motion applied to an element of the shell gives
\begin{equation}
{d\over dt} \left[M (U_0 - U_1)\right]  = 4\pi R^2 P_{\rm int}.
\end{equation}
For a self-similar solution the interior pressure $P_{\rm int}$ will be some fixed fraction $\alpha$ of the post-shock pressure $A(U_0 - U_1)$.  Thus
\begin{equation}
{d\over dt} \left[M(U_0- U_1)\right] = 4\pi R^2 \rho_0 U_0 (U_0-U_1)\alpha
\end{equation}
or, noting that $U_0 = dR/dt$,
\begin{equation}
{R\over U_0-U_1} {d(U_0-U_1)\over dR} = 3(\alpha-1)
\end{equation}
from which it is easy to see that the self-similar Sedov-like solution with $R\propto t^{2/5}$ and $U\propto R^{-3/2}$ requires $\alpha = 1/2$. Thus the interior pressure is half the immediate post-shock pressure.  That there has to be such a pressure gradient in the shell is obvious because the material that was shocked at early times is moving too fast and would overtake the shell were it not decelerated by an adverse pressure gradient directed towards the interior of the remnant. It is however remarkable that the Sedov scaling requires the total interior pressure to be half the ram pressure of the shock,  again with only a very weak dependence on details of equation of state or particle acceleration.  It follows trivially that the total interior energy of the remnant (hot gas and cosmic rays) is between $3/2$ (gas dominated) and $3$ (cosmic ray dominated) times the kinetic energy of the remnant and thus in the two extreme cases we have, if gas dominated,
\begin{eqnarray}
E_{\rm K} &=& {2\over 5} E_{\rm SN},\\
P_{\rm int} &=& {2\over 5} {3 E_{\rm SN}\over 4 \pi R^3},
\end{eqnarray}
and if cosmic ray dominated,
\begin{eqnarray}
E_{\rm K} &=& {1\over 4} E_{\rm SN},\\
P_{\rm int} &=& {1\over 4} {3 E_{\rm SN}\over 4 \pi R^3}.
\end{eqnarray}
Strictly speaking, in the gas dominated case we should not neglect the thickness of the shell, and we should also allow for the fact that the post-shock gas velocity is only $3/4$ of the shock velocity.  Also in the cosmic-ray dominated case, if $\Phi$ has a significant component from genuine escape as distinct from geometrical dilution, this will slightly affect the Sedov exponent.  However the main point of this section is to demonstrate that the interior pressure is rather tightly constrained and is of order $0.4$ to $ 0.25$ times the explosion energy divided by the remnant volume, and that the shock expansion is well approximated by the standard Sedov formula.

Because the total pressure is essentially fixed, if we imagine more and more efficient particle acceleration putting more pressure into $P_C$, the thermal gas pressure $P_G$ has to decrease.  At the same time the nonlinear reaction terms are making the shock more compressive and increasing the downstream density.  Both effect cause the post-shock temperature $T_{\rm i}\propto P_G/\rho$ to decrease so that we expect substantially lower post-shock gas temperatures than in pure gas models of SNRs.
Once a fluid element (in this case a spherical annulus) has been shocked it gradually moves back through the dense shell expanding and dropping in pressure until it merges into the tenuous interior region.  The shell thickness is determined by
\begin{equation}
4\pi R^2 \Delta R \rho_1 \approx {4\pi\over 3} R^3 \rho_0
\end{equation}
or $\Delta R \approx R/ 3 s < R/12$.  The time taken to transit through the shell is thus
\begin{equation}
\Delta t \approx {\Delta R\over U_1} \approx {R\over 3 r} {r\over \dot R} = {1\over 3} {R\over \dot R} ={5\over 6} t
\end{equation}
for a Sedov scaling, close enough to $t$.

Thus if we consider a fluid element that is shocked at time $t_0$, it exits the shell at approximately $(1+5/6) t_0$ with a pressure half the post-shock pressure at that time, which in turn is a factor $(1+5/6)^{-6/5}\approx 1/2$ what it was at the initial time $t_0$.
Thus between $t_0$ and about $2t_0$ the pressure has to drop by a factor of about 4.
After the fluid element has entered the interior region it is a reasonably good approximation to say that it then simply expands in pressure equilibrium with the interior of the remnant where the total pressure is dropping as
\begin{equation}
P_{\rm int} \propto \Esn R^{-3} \propto t^{-6/5}.
\end{equation}
Physically this is because the interior flow is subsonic and any pressure variations get equilibrated on the sound-crossing time which is short compared to the dynamical time of the remnant.  So the history of the parcel of shocked gas is of an initial, rather rapid, drop in pressure by a factor of roughly four between $t_0$ and $2t_0$ followed by a slower power-law $t^{-6/5}$ decline as the remnant expands.

Neglecting for the moment any diffusion of the accelerated particles out of the shell and splitting the total pressure into a ``gas'' component $P_G$ from thermal ions and a ``cosmic ray'' component $P_C$ from the relativistic accelerated particles, we have
\begin{eqnarray}
P_G &\propto& \rho^{5/3}\\
P_C &\propto& \rho^{4/3}
\end{eqnarray}
and thus the gas pressure drops faster than the cosmic ray pressure in the expansion.  It follows that if the two were initially comparable, the cosmic ray pressure will dominate at later stages of the expansion and the density will have to drop according to \begin{eqnarray}
\rho &\propto& t^{-9/10}\\
P_C &\propto& t^{-6/5}\\
P_G &\propto& t^{-3/2}
\end{eqnarray}
and thus the thermal ion temperature will drop as
\begin{equation}
T_{\rm i} \propto {P_G\over\rho} \propto t^{-3/5},
\end{equation}
significantly faster than in a remnant not dominated by cosmic rays where the ion temperature drops as $t^{-12/25}$. This of course ignores the diffusion of cosmic rays  in the interior, but as long as the pressure is more or less uniform this will not affect the result significantly.  The key point is that the cosmic rays come to dominate the pressure as the system expands and this leads to a more rapid cooling of the thermal ions.

In summary therefore the effect of increasing particle acceleration is to leave the rate of expansion of the remnant, and the Sedov scaling with external density and explosion energy, largely unchanged but to make the shock more compressive, the thermal gas colder and the internal pressure become cosmic-ray dominated.

\subsection{Self-regulated acceleration}

The accelerated particle pressure is determined by the interaction of two processes:
\begin{itemize}
\item (i) The first is the injection of particles into the acceleration process at low energies (the injection momentum $p_{\rm inj}$for protons is typically taken as a few times the thermal energy of the shock-heated protons) which determines the number of particles available for acceleration.  There are good reasons for thinking that the injection preferentially favours high rigidity species, but we have no soundly based theory which would allow us to quantitatively predict the injection of various species.  However at least for non-relativistic shocks theoretical considerations and the evidence from heliospheric shocks points to a relatively easy injection of ions as long as the shock is of reasonable strength.

\item (ii) The second is the separation between the injection scale and the upper cut-off of the accelerated particle spectrum.   The key difference between supernova remnant shocks and heliospheric shocks is that in the former the length and time scales are such that a shock-accelerated spectrum can be established which stretches over many decades giving a very large lever-arm between injection and the upper cut-off momentum $p_{\rm max}$.  For the power-law spectra predicted by the linear theory of shock acceleration this large lever arm makes the total pressure very sensitive to the spectral slope.  

\end{itemize}

The pressure per logarithmic interval of momentum is given by the integrand in
\begin{equation}
\int_{p_{\rm inj}}^{p_{\rm max}} {p v\over 3}\, 4\pi p^2 \, f(p) p {dp\over p}
\end{equation}
which scales as $p^5 f(p)$ in the non-relativistic region and as $p^4 f(p)$ in the relativistic region of the spectrum.  Thus if the accelerated particle spectrum is harder than $p^{-5}$ the pressure integrand is a rising function of momentum in the non-relativistic region, and if it is harder than $p^{-4}$ it is a rising function in the relativistic region.  As is well know simple test particle theory gives a slope for the power-law spectrum of accelerated particles which depends only on the compression ratio of the shock,
\begin{equation}
{p\over f} {\partial f\over \partial p} = - q = - {3 s\over s-1}.
\end{equation}
Thus, for a strong shock in a ideal monatomic gas  (i.e. an adiabatic exponent of $5/3$) with compression ratio $s=4$ we get a power-law index of 4, so that the pressure rises linearly with momentum in the non-relativistic region and is then logarithmically divergent in the relativistic region.

The effect of the reaction of the accelerated particles back on the flow is twofold (see \citealt{malkovrev} for a review).
\begin{itemize}
\item (i) It makes the overall shock structure more compressive, and thus for particles that are sufficiently energetic to see this increased shock compression the spectrum hardens with the consequence that the pressure integrand has a maximum close to the upper cut-off.
\item (ii) At low energies however the subshock, where the dissipation is determined by collisionless shock physics operating on plasma scales, is weakened and the associated spectrum of accelerated particles becomes softer.
\end{itemize}
Thus the overall spectrum becomes concave, with a softer portion at low energies becoming harder at high energies and then cutting-off, in an approximately exponential fashion, at some maximum energy.  

A number of approaches have been used to calculate such modified spectra involving various approximations, and the results are in good general agreement (although it must be said that all these approaches make the questionable assumption, explicitly or implicitly, that the shock structure on intermediate scales is quasi-stationary).  The fundamental model behind these solutions is that the shock modification has to adjust itself to throttle back the injection and initial acceleration to avoid an excessive energy demand, and this requires the modification to significantly weaken the subshock compression.  In the limit that that injection energy can be treated as  small compared to the proton rest mass energy, so that there is a long lever-arm between injection and the relativistic regime, the above arguments suggest that the spectrum in the non-relativistic region should be close to $p^{-5}$ and thus the subshock compression should approach the critical value of $s_{\rm sub}\approx 2.5$. 

If the subshock compression is weaker than 2.5 the pressure integrand can have no local minimum separating a ``thermal'' population from a ``non-thermal'' component and instead there is just a very weak non-thermal tail on a dominant thermal population.  On the other hand, if the subshock is stronger than compression 2.5, the pressure integrand will start to rise, there will be a minimum in the pressure integrand, and we can usefully talk about two particle populations; a bulk ``thermal'' component and a non-thermal component where the energy is concentrated in relativistic particles.

This simple picture is confirmed by the nonlinear models, see e.g.\ the discussion in \citet{BE99} and in particular their figure 6(b), which show that for efficient injection and strong shocks the subshock compression is typically between 2 and 3 with a rather weak dependence on total Mach number.  The recent study \citep{KRJ}
also finds an asymptotic subshock compression of $3.2(M/10)^{0.04}$ for $M>10$ where $M$ is the shock Mach number.  

\section{Ion  temperatures}

If the subshock compression automatically adjusts to a value close to 2.5 to 3, it follows that the sub-shock Mach number, and the amount of gas heating in the subshock, is essentially fixed. {\em This has the very  important consequence that the downstream gas temperature is determined largely by the upstream gas temperature and not, as in standard shock models, by the shock speed.}  The inflowing gas is first heated by adiabatic compression in the shock precursor from the far upstream value $T_0$ to a value $T_1$ just in front of the subshock of
\begin{equation}
T_1 = T_0 s_{\rm pre}^{\gamma-1}
\end{equation}
where $s_{\rm pre}$ is the precursor compression
and then on passage through the compression 2.5 subshock by an amount
\begin{equation}
T_2 = {12\over 5} T_1
\end{equation}
(assuming a $\gamma=5/3$ polytropic equation of state for the gas).
This rises to 
\begin{equation}
T_2 = {11\over 3}T_1
\end{equation}
for a subshock of compression ratio 3, still a relatively modest increase (if there is significant wave dissipation in the precursor it is of course possible to raise the gas temperature to higher levels \citep{VDM}).

If the total compression, as suggested by numerous simulation studies, is of order $10$,
with a factor of $4$ in the precursor and $2.5$ in the subshock it follows that the downstream temperature is just the far upstream temperature multiplied by a factor of 
\begin{equation}
4^{2/3} {12\over 5} \approx 6.05
\end{equation}
and is independent of the shock speed.  This is in marked contrast to the standard picture where the downstream temperature is determined purely by the shock speed (with a strong quadratic dependence)  and is only very weakly dependent on the upstream temperature according to,
\begin{equation}
{3\over 2} kT_G \approx  {1\over 2} m_p \left({3\over 4} U\right)^2,
\end{equation}
$k$ being Boltzmann's constant and $m_p$ the proton mass.  In reality the total compression does increase with shock Mach number (\citealp{KRJ} suggest roughly as the one third power, although this calculation includes Alfven wave heating effects;  a naive estimate without wave heating suggests that the exponent might be as high as 3/4) and thus there is a weak dependence of the post-shock temperature on shock speed (scaling somewhere between 2/9 and1/2 power) but nothing as strong as in the unmodified case.

Clearly this is an extreme limit, but the rather surprising answer to the question, how cold can the postshock gas be in the limit of strong particle acceleration, is as cold as six times the upstream temperature!  

Motivated by the above discussion we now give a simple description of the gas compression and heating in a supernova remnant with strong particle acceleration.  As an illustrative example we follow a fluid element as it is swept up by the shock and compressed, initially to a density of 10 times ambient and temperature of order 6 times ambient at remnant age $t_0$.  We assume that it then expands and cools so that the total pressure drops initially by a factor of order 4 over the time $t_0 <t < 2t_0$ and then in pressure equilibrium with the interior.  Consistent with the assumption of strong acceleration we assume that the cosmic ray pressure dominates the thermal gas pressure (note that  if this is true initially, it is true at all later times).  In this situation we have $P_C\propto\rho^{4/3}$ and $T_G \propto P_G/\rho \propto \rho^{2/3}$ so that $T_G\propto P_C^{1/2}$.

\begin{figure}[htbp]
\begin{center}
\includegraphics[width=\hsize]{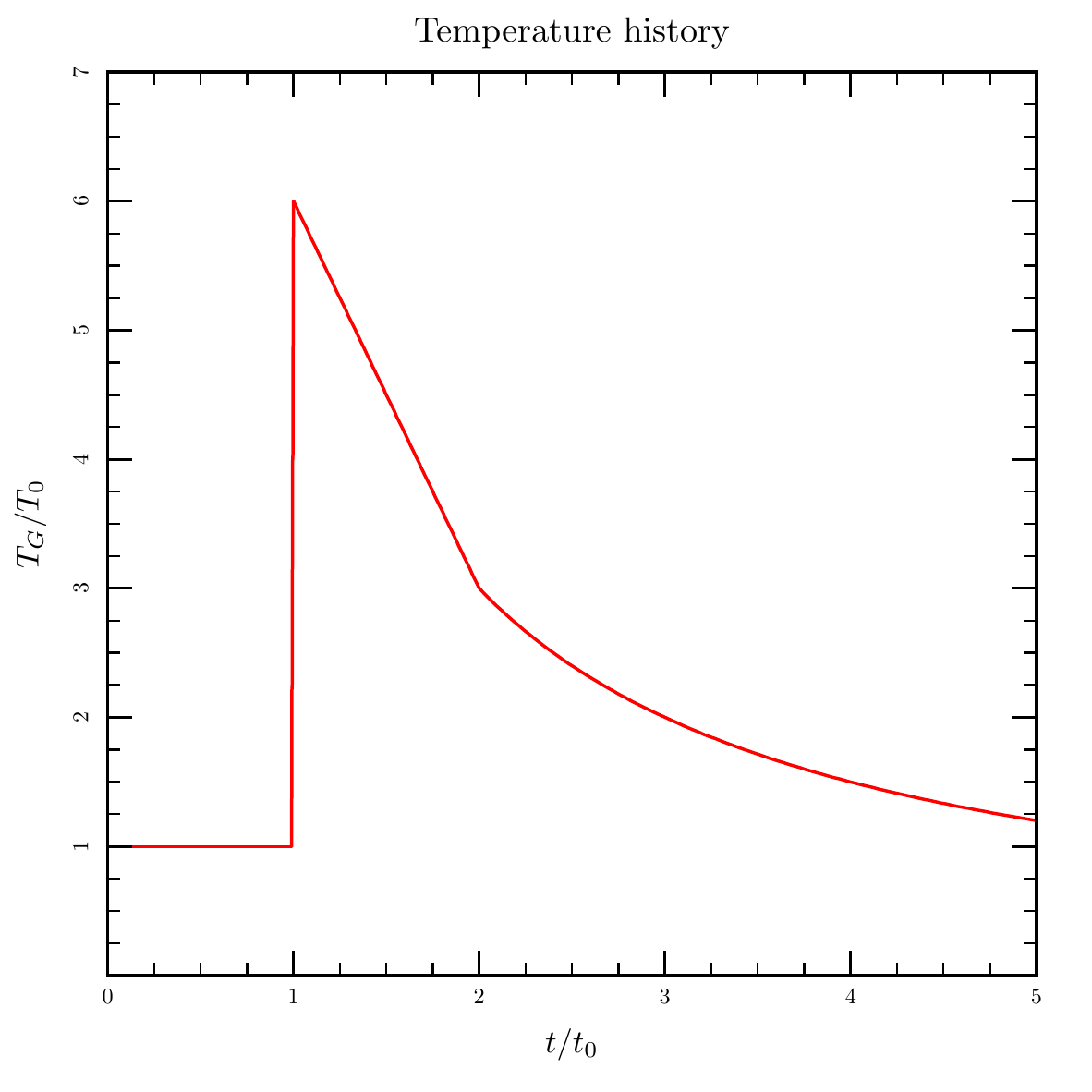}
\caption{The approximate temperature history as a function of time $t$ of an element of gas shocked at time $t_0$.  The graph shows a sketch of the plasma temperature, $T_G$, in terms of the far upstream ambient temperature $T_0$ for case of  no wave heating in the precursor, strong cosmic ray acceleration and an assumed total compression of 10.}
\label{TempHist}
\end{center}
\end{figure}

Thus the gas temperature can be approximated as
\begin{eqnarray}
{T_G \over T_0} &\approx & 6 - 3\left({t\over t_0} -1\right) \qquad t_0 < t < 2t_0\\
&\approx& 3 \left(t\over 2 t_0\right)^{-3/5} \qquad t> 2 t_0
\end{eqnarray}
and the density as
\begin{equation}
{\rho\over\rho_0}\approx 10 \left(T_G\over 6 T_0\right)^{3/2}
\end{equation}
This is illustrated in Fig.~\ref{TempHist}.

Even assuming full thermal equilibrium between the electrons and the ions, it is clear that the gas temperature can be reduced to the point where thermal X-ray emission is severely suppressed unless the upstream medium is itself already very hot.  It is interesting in this context to note the observational suggestion for substantially lower than expected ion and electron temperatures seen in the Magellanic cloud remnant 1E 0102.2-7219 \citep{HRD}. 

\section{Conclusions}

We have attempted to outline the physics behind the modifications of SNR structure and dynamics by particle acceleration.  {\em While the overall dynamics are relatively insensitive, we have shown that the gas heating can in principle be strongly suppressed to the point where thermal X-ray emission is no longer expected}.  This has important consequences for the on-going debate on the origin of the TeV emission detected from a number of young shell-type SNRs.

\appendix
\section{The accelerated particle flux}

The expression (9) for the number flux of particles being accelerated upwards in momentum at the shock can be derived in a number of different ways and is fundamental to the theory of shock acceleration.

Perhaps one of the more physical derivations which emphasises the generality of the result is 
the following.  We begin by considering a general oblique non-relativistic MHD shock in the standard de Hoffman-Teller frame where the plasma velocity is always parallel to the magnetic field and the electric field vanishes identically.  Let the plasma velocity be $\vec U_1$ upstream and $\vec U_2$ downstream
with both velocities small relative to the speed of light $c$ and let the shock normal direction  be $\vec n$.

Let us now consider a relativistic particle of momentum $\vec p$ and velocity $\vec v$ sufficiently energetic that its gyroradius is much larger than the shock thickness. In the de Hoffman-Teller frame it simply crosses the shock front ballistically with no change to momentum or velocity.  However we conventionally measure particle energies relative to a frame co-moving with the local plasma, and in these reference frames there is a change.  Transforming into the local plasma frame from the de Hoffman-Teller frame induces a first-order change in the magnitude of the particles momentum of
\begin{equation}
{\Delta p} =  {\vec p \cdot \vec U\over v}
\end{equation}
as can be trivially verified by carrying out the Lorentz transformation of the particle's energy momentum 4-vector and neglecting terms of second order and higher in $U/c$.  There is thus a change in momentum, measured in the local co-moving plasma, of
\begin{equation}
{\Delta p} =  {\vec p \cdot (\vec U_1 - \vec U_2)\over v}
\end{equation}
for the particle crossing the shock.

Let us now consider a distribution of particles incident with different velocity vector directions on the shock.   If we consider velocity vectors within a certain solid angle $d\Omega$ the flux across the shock front is 
\begin{equation}
p^2 f(\vec p)\, \vec v \cdot \vec n \,d\Omega
\end{equation}
and thus the flux of particles upwards through momentum level $p$ at the shock is 
\begin{equation}
\Psi(p) = \int p^2 f(\vec p)\, \vec v \cdot \vec n\, {\vec p \cdot (\vec U_1 -  \vec U_2)\over v}\,d\Omega.
\end{equation}
In general this is all that one can say, but if the distribution function is close to isotropic there is a further significant simplification.  In the special case of isotropic distributions, by symmetry the integral must reduce to a scalar times the dot product  $\vec n  \cdot (\vec U_1 - \vec U_2)$ and by evaluating in the special case of a parallel shock (with $\theta$ the angle between the particle velocity vector and the shock normal) we easily find
\begin{eqnarray}
\Psi(p) &=& p^3f(p) \int \vec n \cdot (\vec U_1 - \vec U_2) \cos^2(\theta) d\Omega \nonumber\\
&=& {4\pi \over 3}p^3f(p) \vec n \cdot (\vec U_1 - \vec U_2)
\end{eqnarray}
as in equation (9).

This emphasises that the result depends only on the distribution function of the accelerated particles being close to isotropy at the shock and the velocity shifts into the local plasma frame being sub-relativistic; the two assumptions are of course closely related the main problem in the theory of relativistic shock acceleration being that the distributions are highly anisotropic at the shock.

Another, perhaps more conventional, route to this result is to write the basic diffusive transport equation in conservation form.  Normally the transport equation is written in the form
\begin{equation}
{\partial f\over\partial t} + \vec U \cdot \nabla f 
= {1\over 3} \nabla\cdot \vec U p{\partial f\over \partial p} 
+\nabla\left(\kappa\nabla f\right),
\end{equation}
but we can equivalently write this as
\begin{eqnarray}
{\partial\over\partial t}\left(4\pi p^2 f(p)\right) 
&+&\nabla\cdot \left(4 \pi p^2 f(p) \vec U - 4 \pi p^2 \kappa \nabla f\right)\nonumber\\
&+&{\partial\over\partial p}\left[{4\pi p^3\over 3} f \left(-\nabla\cdot \vec U\right)\right]
=0
\end{eqnarray}
which can be interpreted as stating that the time rate of change of the number density, 
plus the physical space divergence of the advective and diffusive fluxes, plus the momentum space divergence of the acceleration flux is zero.  Thus the acceleration flux in momentum space is
\begin{equation}
\Psi = {4\pi p^3\over 3} f(p) \left(-\nabla\cdot \vec U\right)
\end{equation} 
and is directly proportional to the flow compression.  In the case of a shock the compression is concentrated in a Dirac delta distribution at the shock itself
\begin{equation}
-\nabla\cdot \vec U = \vec n \cdot \left(\vec U_1 - \vec U_2\right) \delta (x)
\end{equation}
and thus we get a localised acceleration flux at the shock,
\begin{equation}
\Psi = {4\pi p^3\over 3} f(p)  \vec n \cdot \left(\vec U_1 - \vec U_2\right) \delta (x)
\end{equation} 
in agreement with equation (9).


\begin{thebibliography}{99}


\bibitem[Aharonian et al.(2007)]{HESS_last} Aharonian, F.A. et al. 2007 (HESS collaboration),  A\&A, 464, 235

\bibitem[Bell(2004)]{bell} Bell, A.R. 2004, MNRAS, 353, 550

\bibitem[Berezhko and Ellison (1999)]{BE99} Berezhko, E. G. and Ellison, D. C., Ap. J., 526, 385 

\bibitem[Blasi et al. (2005)]{giulia} Blasi, P., Gabici, S., Vannoni, G. 2005, MNRAS, 361, 907

\bibitem[Cassam-Chenai et al. (2004)]{XMM} Cassam-Chenai ,G., Decourchelle, A.,
Ballet, J.,  Sauvageot, J.-L., Dubner, G.,  Giacani, E. 2004, A\&A, 427, 199

\bibitem[Chernyi(1957)]{Chernyi} Chernyi, G. G. 1957, Dokl. Akad. Nauk SSSR, 112, 213

\bibitem[Drury and V\"olk (1981)]{DV81} Drury, L. O'C. and V\"olk, H. J., Ap. J  248 (1981) 344--351

\bibitem[Drury et al. (1989)]{DMV} Drury, L.O'C., Markiewicz, W. J., and V\"olk, H. J., A\&A  225 (1989) 179--191

\bibitem[Ellison(1985)]{ellison85} Ellison, D. C., JGR 90 (1985) 29--38.

\bibitem[Ellison(2000)]{don} Ellison, D.C., astro-ph/0003214

\bibitem[Hughes et al. (2000)]{HRD} Hughes, J. P., Rakowski, C. E. \& Decourchelle, A. Ap.J 543 (2000) L61-65.

\bibitem[Gabici(2008)]{Gabici} Gabici, S. 2008, arXiv:0811.0836 

\bibitem[Kang et al. (2008)]{KRJ} Kang, H., Ryu, D., \& Jones, T. W., 2008 personal communication.

\bibitem[Katz and Waxman (2008)]{Katz} Katz, B. and Waxman, E. 2008, JCAP,  01, 018.

\bibitem[Malkov (1999)]{malkov}
Malkov, M.A. 1999, ApJ, 511, L53

\bibitem[Malkov and Drury (2001)]{malkovrev}
Malkov, M.A., Drury, L.O'C. 2001, Rep. Prog. Phys., 64, 429

\bibitem[Reville, Kirk and Duffy(2008)]{RKD}Reville, B., Kirk, J. G. and Duffy, P. 2008
arXiv:0812.3993

\bibitem[Slane et al. (1999)]{Slane} Slane, P., Gaensler, B.,  Dame, T. M.,  Hughes, J.P., Plucinsky, P. P.,
Green, A. 1999,  ApJ, 525, 357

\bibitem[Takahashi et al. (2008)]{Suzaku}  Takahashi, T. et al. 2008, PASJ, 60, S131

\bibitem[Tanaka et al. (2008)]{Tanaka}  Tanaka, T. et al. 2008, ApJ, 685, 988

\bibitem[V\"olk et al. (1984)]{VDM} Voelk, H.J., Drury, L. O'C. \& McKenzie, J. F. A\&A 130 (1984) 19--28.

\bibitem[Zel'dovich and Raiser (1966)]{Zeld} Zel'dovich, Ya. B.  and  Raiser, Yu., P. 1966, 
``Physics of Shock Waves and High-temperature Phenomena'', Academic Press, New York





\end{thebibliography}
\end{document}